# Realization of $\frac{2}{3}$-layer transition metal dichalcogenides

Ya-Xin Zhao[1,3,†], Zi-Yi Han[1,3,†], Ya-Ning Ren[1,3], Ruo-Han Zhang[1,3], Xiao-Feng Zhou[1,3], Yu Zhang[2,*], and Lin He[1,3,*]

[1]Center for Advanced Quantum Studies, Department of Physics, Beijing Normal University, Beijing, 100875, People's Republic of China.

[2]Advanced Research Institute of Multidisciplinary Sciences, Beijing Institute of Technology, Beijing 100081, China.

[3]Key Laboratory of Multiscale Spin Physics, Ministry of Education, Beijing, 100875, China

[†]These authors contributed equally to this work.

[*]Correspondence and requests for materials should be addressed to Yu Zhang (e-mail: yzhang@bit.edu.cn) and Lin He (e-mail: helin@bnu.edu.cn).

**Layered van der Waals transition metal dichalcogenides (TMDCs), generally composed of three atomic X-M-X planes in each layer (M = transition metal, X = chalcogen), provide versatile platforms for exploring diverse quantum phenomena. In each $MX_2$ layer, the M-X bonds are predominantly covalent in nature, as a result, the cleavage of TMDC crystals always occurring between the layers. Here we report the controllable realization of fractional-layer $WTe_2$ via an in-situ scanning tunnelling microscopy (STM) tip manipulation technique. By applying STM tip pulses, hundreds of the topmost Te atoms are removed to form a nanoscale monolayer Te pit in the 1T′-$WTe_2$, thus realizing a brand-new $\frac{2}{3}$-layer $WTe_2$. Such a unique configuration undergoes a spontaneous atomic reconstruction, yielding an energy-dependent unidirectional charge-density-wave state with the wavevector and geometry quite distinct from that of pristine 1T′-$WTe_2$. Our results expand the conventional understanding of the TMDCs and are expected to stimulate the research on extraordinary structures and properties based on fractional-layer TMDCs.**

Layered transition metal dichalcogenide (TMDC) materials have recently attracted remarkable interest, owing to their intriguing physical properties of superconductivity and topology[1-3] as well as broad applications in quantum devices and information technologies[4-7]. Generally, TMDCs have a $MX_2$ formula and, within each TMDC layer, a plane constructed of transition metal atoms (M) is sandwiched between two planes of chalcogen atoms (X). The intralayer X-M-X configuration determines the versatile structural phases of TMDC, including trigonal prismatic (2H), octahedral (1T), and distorted octahedral (1T′) phases[1,2,8]. Since the intralayer M-X bonds are predominantly covalent in nature, much stronger than the interlayer van der Waals (vdW) forces, as a sequence, TMDCs exhibit a quasi-2D form with the intralayer stability. Indeed, traditional cleavage of TMDC crystals always occurs between the layers[9-12].

The realization of fractional-layer TMDCs by removing an atomic chalcogen plane or two atomic chalcogen-metal planes has long thought to be impossible in experiments. Although chalcogen-atom vacancies in TMDCs are ubiquitous, they are always discretely distributed and can hardly be aggregated within a two-dimensional (2D) region to form a monolayer chalcogen pit, because high-density dangling bonds of the exposed transition-metal atoms are chemically active and extremely sensitive to the environments[13-17]. Contradicting this belief, here we successfully fabricate fractional-layer $WTe_2$ via an in-situ scanning tunnelling microscopy (STM) tip manipulation technique. By applying a STM tip pulse onto graphene-covered 1T′-$WTe_2$ crystals, hundreds of the topmost Te atoms are simultaneously removed to form a monolayer Te pit in the 1T′-$WTe_2$, yielding a brand-new $\frac{2}{3}$-layer $WTe_2$ with the lateral size larger than 10 nm. Our high-resolution topographic and spectroscopic measurements further reveal the existence of spontaneous atomic reconstruction and unidirectional charge-density-wave (CDW) order in the $\frac{2}{3}$-layer $WTe_2$, which have never been reported in pristine 1T′-$WTe_2$. These results open an avenue to explore fractional-layer TMDCs and their heterostructures with exotic properties beyond their integer-layer components.

In our experiments, high-quality 1T′-WTe$_2$ multilayers were prepared onto silicon substrates and covered by graphene monolayer, as schematically shown in Fig. 1a (see Methods for details). The 1T′ phase of WTe$_2$ is originated from the spontaneous lattice distortion of its 1T phase where the W atoms show a lateral distortion towards the $b_1$ direction. This creates one-dimensional (1D) zigzag chains along the $a_1$ direction with the lattice constant $a_1$ = 3.49 Å and $b_1$ = 6.65 Å[15,21,22]. Such a W-atom distortion further drives the bilateral Te atoms exhibiting an out-of-plane distortion, yielding the neighboring Te-atom chains with different heights[18-22]. The 1T′-WTe$_2$ structure can be easily identified in atomic-resolution STM topography, which mainly reflects the information of the topmost Te atoms, by the doubling periodicity of the alternating bright and dark chain contrasts, as exhibited in Fig. 1b.

STM tip pulse has been widely regarded as a powerful means of manipulating local structures of materials with nanoscale precision[23-25]. In our experiments, after applying a STM tip pulse over the threshold of 4 V onto graphene-covered 1T′-WTe$_2$, there is usually a nanoscale pit appearing beneath the tip, as typically shown in Fig. 1c-e of the STM images and the corresponding height profiles across the pits. Within the pits, the atomic topography usually exhibits an obvious periodicity (Fig. 1e), with the lattice constants and orientations quite different from those of the outside region, highlighting a newly emerging atomic configuration evolved from the 1T′-WTe$_2$. Similar phenomenon has been realized in tens of graphene-covered 1T′-WTe$_2$ samples by using different STM tips (Supplementary Figs.1 and 2), which helps us rule out any possible artifacts. Such an atomic configuration can be identified as a $\frac{2}{3}$-layer WTe$_2$ that the topmost Te-atom plane of the 1T′-WTe$_2$ is removed, as we will demonstrate below.

At the atomic limit, a STM tip pulse can remove an individual Te atom in the topmost 1T′-WTe$_2$ layer. The individual Te vacancy can be clearly distinguished via the STM topography by superposing the atomic structure onto it (Fig. 1c and Supplementary Fig. 3), as also observed in previous studies[19,22,26]. As the lateral size of the tip-induced pit increasing, the outline of the pit varies from an atomic void (Fig. 1c) to a nanoscale hexagon with a set of opposite edges parallel to the $a_1$ direction of the 1T′-WTe$_2$ (Fig. 1e). Simultaneously, the depth of the pit gradually increases and finally reaches the maximum of about 278 pm when the lateral size exceeds 10 nm, as summarized in Fig. 1f. It's worth noting that the maximum depth of the pits is quite consistent with the height of an individual Te-atom plane in the 1T′-WTe$_2$[26]. The hexagonal structure and the measured depth of the pit indicate that the pit is generated by removing hundreds of the topmost Te atoms in the 1T′-WTe$_2$. Because of the covered graphene monolayer, these released Te atoms generated by breaking the intralayer W-Te bonds usually aggregate at one corner of the pit, showing as a protrusion in the STM image (Fig. 1e).

The removing of the topmost Te atomic plane to form the unique $\frac{2}{3}$-layer WTe$_2$ is further confirmed by measuring the formation process of the hexagonal pits. Figure 2a shows a representative example. In our experiment, in a few cases, we can observe tip-pulse-induced irregular hexagonal pit, as typically shown in Fig. 2a, that is constructed of two straight edges and a half arc approximately. Our experiment indicates that this configuration is not stable during the STM measurements and the half-arc outline of the pit expands outward and evolves into four straight edges, yielding a regular hexagon as a whole (Fig. 2b). In the meanwhile, more Te atoms from the pit are aggregating, resulting in an increasing area of the protrusion, as schematically shown in Fig. 2c,d. Taken together, our experiments undoubtedly demonstrate the formation of the $\frac{2}{3}$-layer WTe$_2$ by breaking the intralayer W-Te covalent bonds between the topmost W and Te atomic planes under a STM tip pulse.

To shed light on the formation mechanism of the $\frac{2}{3}$-layer WTe$_2$, firstly, we carried out the STM measurements by applying tip pulses onto a bared 1T′-WTe$_2$, i.e., without covering graphene. Surprisingly, there is no signature of the $\frac{2}{3}$-layer WTe$_2$, instead, the depth of tip-induced pits is usually of several nanometers (see Supplementary Fig. 4). This observation indicates that the covered graphene monolayer helps to stabilize the $\frac{2}{3}$-layer WTe$_2$ configuration, which is reasonable since graphene hosts the ability of protecting the covered materials from oxidation[27]. We also carried out the same experiments on graphene-covered 2H-WSe$_2$ and 2H-MoTe$_2$ for comparison. From the STM images given in Supplementary Fig. 5, we can find out the depths of the tip-induced pits are about 800 nm, in consistent with a monolayer thickness, highlighting the absence of fractional-layer components in these systems[28,29]. Previous experiments also reported the realization of a similar monolayer pits in 1T-TiSe$_2$ after the annealing[30]. Considering that a chalcogen vacancy in the 1T′ phase of TMDCs always hosts a smaller formation energy and is more likely to be delocalized than those of the 1H and 1T phases, as well as the vacancy formation energy of the Te-based TMDCs is lower than that of the Se or S compounds[31,32], therefore, the 1T′-WTe$_2$ is expected to be an ideal candidate for realizing a fractional-layer configuration by locally removing an atomic Te sheet.

To further explore the structures of the $\frac{2}{3}$-layer WTe$_2$, we carried out atomic resolved STM measurements. Figure 3a,b shows typical STM images recorded around a $\frac{2}{3}$-layer WTe$_2$ pit under the sample biases of 0.8 V and 1.5 V. The corresponding Fourier transforms are given in Fig. 3c,d, where the bright spots marked by the black, blue, and red circles indicate the reciprocal lattices of the covered graphene monolayer, the 1T′-WTe$_2$ outside the pit, and the $\frac{2}{3}$-layer WTe$_2$ within the pit, respectively. From Fig. 3a-d, we can find out that the wavevector and geometry of the $\frac{2}{3}$-layer WTe$_2$ are strongly dependent on the sample bias, in stark contrast to those of the pristine 1T′-WTe$_2$. This phenomenon highlights that the $\frac{2}{3}$-layer WTe$_2$ undergoes a remarkable lattice reconstruction. Figure 3g-l shows high-resolution STM images and the corresponding Fourier transforms recorded around the pit, as the precise locations marked in Fig. 3a. There are three notable features. Firstly, the Fourier peaks of the graphene lattice are constant around the pit before and after applying a tip pulse (black circles in Fig. 3j-l and Supplementary Fig. 6), implying the covered graphene monolayer is intact under a tip pulse[33] (indeed, the C-C bond energy in graphene is over 7 eV[34]). This result undoubtedly verifies that the tip-induced pit configuration is attributed to the topmost 1T′-WTe$_2$ monolayer and its derivative. Secondly, the released

Te atoms induced by the tip pulse prefer to aggregate and generate a 2D rectangular lattice, analogous to the previous reported β-tellurene[35,36], with the lattice constant of $a_3$ = 4.53 Å and $b_3$ = 5.03 Å and the height of about 0.3 nm (Fig. 3i,l and Supplementary Fig. 7). This phenomenon further confirms that the STM tip pulse can efficiently remove the topmost Te atoms in the 1T′-WTe$_2$, thus realizing the β-tellurene and the $\frac{2}{3}$-layer WTe$_2$ simultaneously.

Thirdly but most significantly, the atomic structure inside the pit shows an obvious periodicity with the lattice constant of $a_2$ = 7.2 Å, $b_2$ = 13.0 Å and an intersection angle of about 90° (Figs. 3h,k). In addition, $a_2$ shows an approximate 60° rotation relative to $a_1$ of pristine 1T′-WTe$_2$. Moreover, the combination of $a_1$, $a_2$, $a_2 + b_2$ directions contributes to the edge orientations of the hexagonal pits, implying the atomic structures outside and inside the pits share the same crystal symmetry at the atomic scale. It's worth noting that all the characteristics are robust regardless of the orientation between graphene and 1T′-WTe$_2$ (see Supplementary Fig. 8), which help us to eliminate the graphene-constructed moiré superlattice as the cause of our observations. Assuming that only the topmost Te atoms of 1T′-WTe$_2$ surface are released under a tip pulse while the underlying W-Te atomic planes are continuous, the exposed W-atom plane maintains a doubling periodicity and exhibits an elliptical shape in each supercell naturally, as schematically shown in Fig. 3e,f. The superlattice inside the pit, as a sequence, can be regarded as the W-atom plane of the $\frac{2}{3}$-layer WTe$_2$ undergoing a spontaneous atomic reconstruction with a fourfold periodicity along the $a_1$ direction (marked by the dark-green ellipses in Fig. 3e), based on our observations (Fig. 3h).

Such a brand-new fractional-layer TMDC provides an unprecedented platform to explore novel electronic states. Figure 4b,c shows spatially resolved scanning tunneling spectroscopy (STS) spectra recorded across the pit along the yellow and blue arrows marked in Fig. 4a. On the graphene-covered 1T′-WTe$_2$ away from the pit, there is a peak appearing at the energy of about 0.6 eV, in consistent with that observed in the 1T′-WTe$_2$[22,37-39]. As approaching the edge of the pit, such a peak gradually becomes weak and finally vanishes. In contrast, the STS spectra recorded on the graphene-covered $\frac{2}{3}$-layer WTe$_2$ show no resonance peak, while exhibit a significant enhancement of the density of states (DOS) in the energy range of -0.5 eV < *E* < 0 eV and 0.7 eV < *E* < 1.0 eV, as clearly shown in Fig. 4d.

Figure 4e-h shows the spectroscopic maps recorded at the same location of Fig. 4a, which can directly reflect the real-space charge distributions at different energies. As we can see, the $\frac{2}{3}$-layer WTe$_2$ always exhibit periodic stripes along the *a$_2$* direction with a period of about 1.2 nm, even though the contrast is strongly dependent on the recorded energy. Specifically, the spectroscopic map exhibits an enhanced striped intensity that follows the STM topography (Fig. 4a) when -0.6 eV < *E* < -0.3 eV (Fig. 4e). As a contrast, it shows an evident anti-phase spatial modulation when -0.3 eV < *E* < 1.0 eV. The periodic charge modulation as well as its anti-phase relation hallmark a CDW state[40-43]. Very recently, similar CDW superlattices with fourfold stripes were also reported in VSe$_2$, VTe$_2$, and NbTe$_2$ systems[44-48]. In our experiment, there is no sign of a CDW gap in the STS spectra of the $\frac{2}{3}$-layer WTe$_2$ (Fig. 4d), which can be partly attributed to the covered graphene that covers up the CDW gap underneath it, and can be partly attributed to the thermal broadening of the spectra since they are carried out at the liquid nitrogen temperature. These results demonstrate that the $\frac{2}{3}$-layer WTe$_2$ hosts an intrinsic striped CDW state, which has never been realized in the 1T′-WTe$_2$.

In summary, we controllably construct the $\frac{2}{3}$-layer WTe$_2$ from the 1T′-WTe$_2$ for the first time. By applying an in-situ STM tip pulse, hundreds of the topmost Te atoms can be removed from 1T′-WTe$_2$ surface, thus realizing the $\frac{2}{3}$-layer WTe$_2$. Such a configuration prefers to undergo a spontaneous atomic reconstruction and generate a unidirectional CDW state, which have never been reported in pristine 1T′-WTe$_2$. Our results open an avenue to explore fractional-layer TMDCs and their heterostructures with exotic properties beyond their integer-layer components.

**Acknowledgements**

This work was supported by the National Key R&D Program of China (Grant Nos. 2022YFA1402502, 2021YFA1401900, 2021YFA1400100, 2022YFA1402602), National Natural Science Foundation of China (Grant Nos. 12141401, 12274026), the Fundamental Research Funds for the Central Universities, and the China Postdoctoral Science Foundation (Grant Nos. 2021M700407, 2023M740296).

# Figures

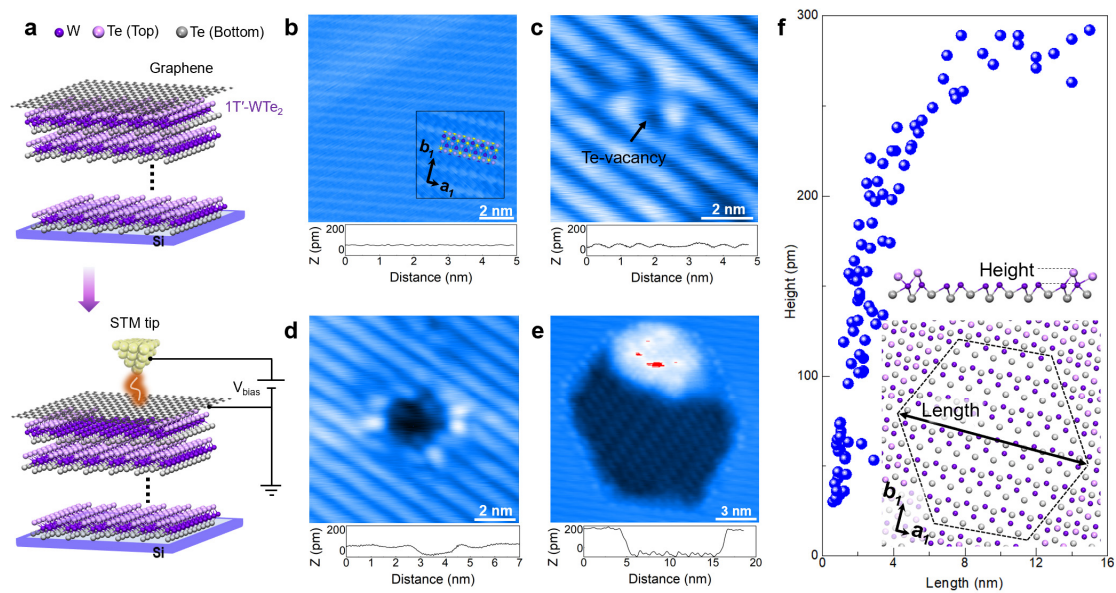

**Figure 1. Nanoscale fabrication of $\frac{2}{3}$-layer WTe$_2$ via a STM tip manipulation technique**. **a,** Experimental set-up. Upper panel: Bulk 1T′-WTe$_2$ crystal is firstly placed on top of Si substrates, and then covered by graphene monolayer. Bottom panel: the realization of $\frac{2}{3}$-layer WTe$_2$ by applying a STM tip pulse. **b,** Large-scale STM image of graphene-covered 1T′-WTe$_2$ before applying a STM tip pulse. Inset: atomically resolved STM image of graphene-covered 1T′-WTe$_2$, with the atomic structures of 1T′-WTe$_2$ superposing on the image. **c-e,** Typical STM images and the corresponding height profiles of graphene-covered 1T′-WTe$_2$ after applying a STM tip pulse. There are obvious pits exhibited in the STM images. **f,** The depth of the tip-induced pit as a function of its lateral size.

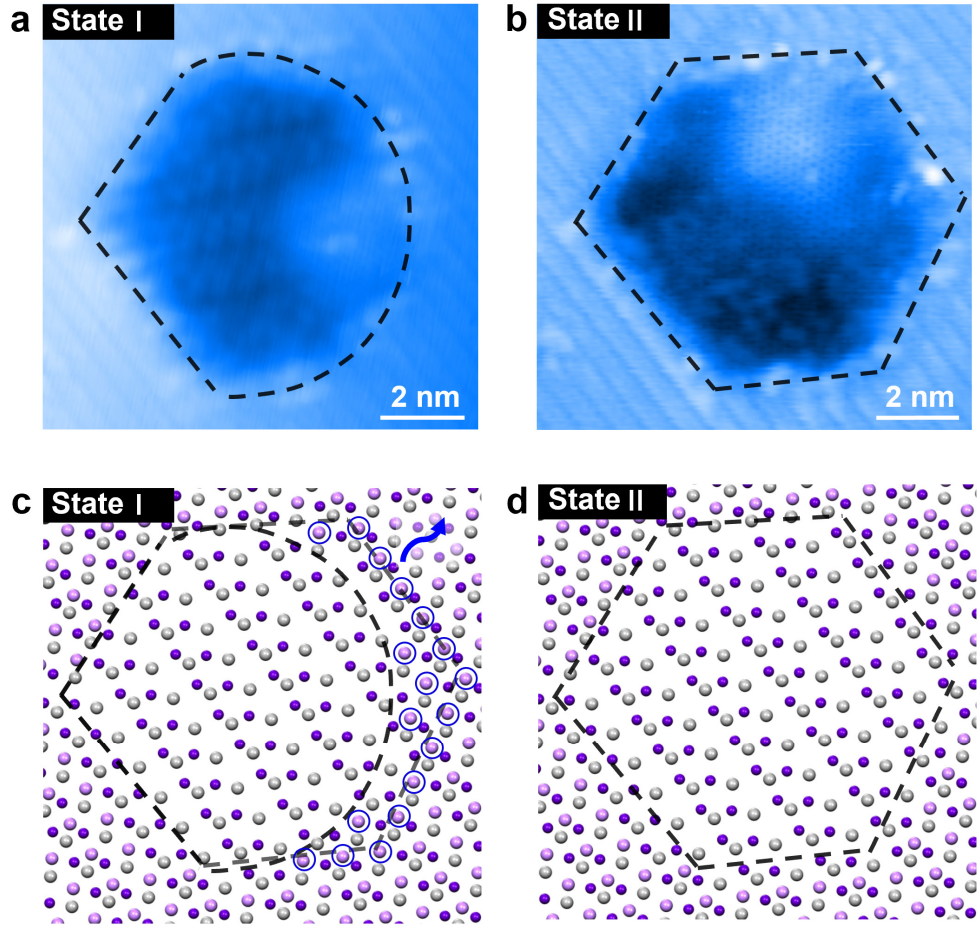

**Figure 2. STM verification for realizing fractional-layer 1T′-WTe$_2$. a,** Typical STM image of graphene-covered 1T′-WTe$_2$ after applying a STM tip pulse. The pit shows an irregular hexagonal outline constructed of two straight edges and a half arc approximately. **b,** Typical STM image after applying an additional STM tip pulse. The half-arc outline of the pit expands outward and evolves into four straight edges, yielding a regular hexagon as a whole. **c,d,** Schematic of the formation process of $\frac{2}{3}$-layer WTe$_2$.

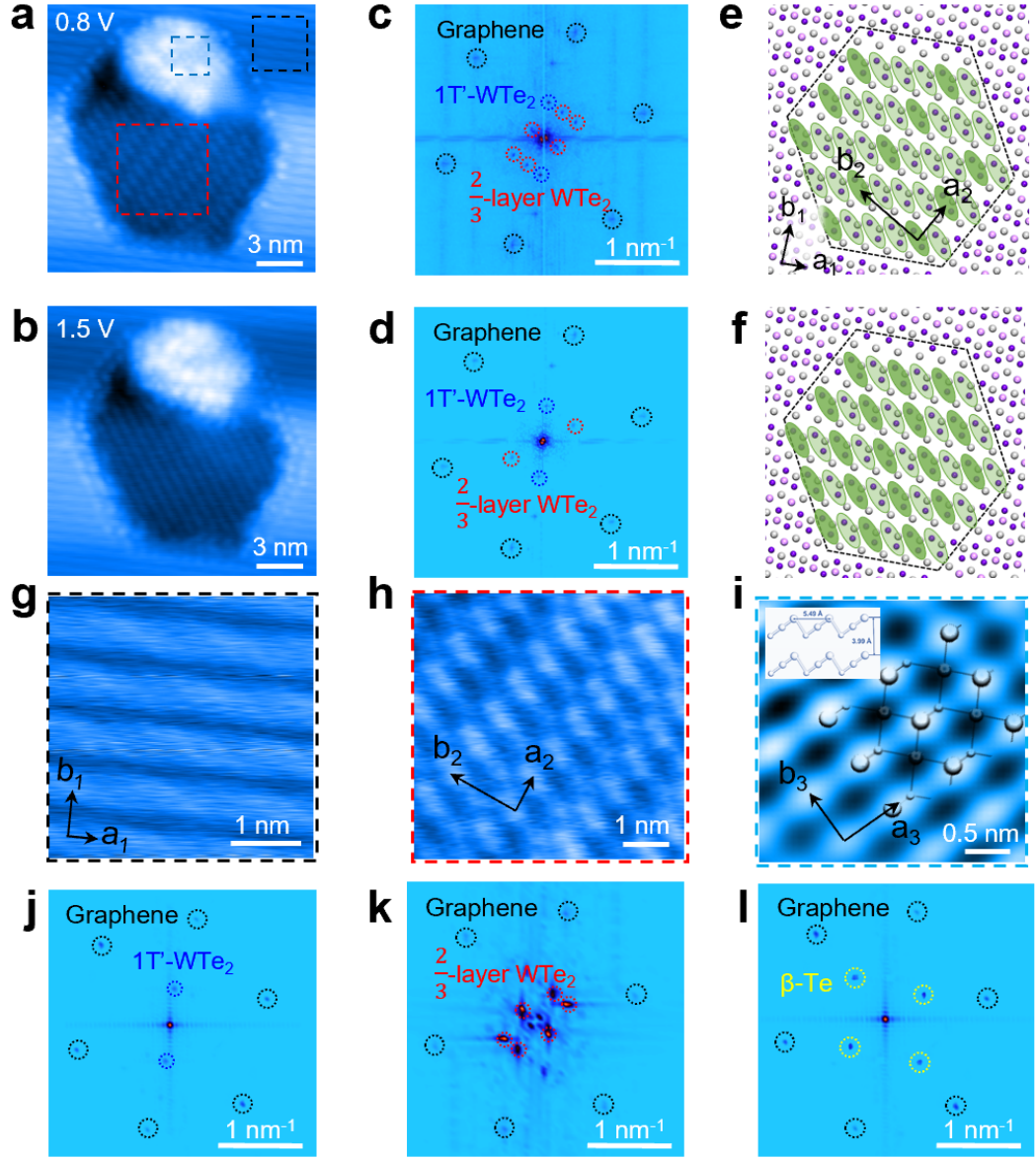

**Figure 3. Atomically resolved STM images around the tip-induced pit. a,b,** Typical STM images of graphene-covered 1T′-WTe$_2$ after applying a STM tip pulse under the sample biases of 0.8 V and 1.5 V, respectively. **c,d,** Corresponding Fourier transforms of panels a and b. The bright spots marked by the black, blue, and red circles indicate the reciprocal lattices of the covered graphene monolayer, 1T′-WTe$_2$ outside the pit, and $\frac{2}{3}$-layer WTe$_2$ within the pit, respectively. **e,f,** Atomic models of panels a and b. **g-i,** Atomically resolved STM images of graphene-covered 1T′-WTe$_2$, $\frac{2}{3}$-layer WTe$_2$, and β-Te recorded at the regions marked in panel a. **j-l,** Corresponding Fourier transforms of panels g-i, respectively.

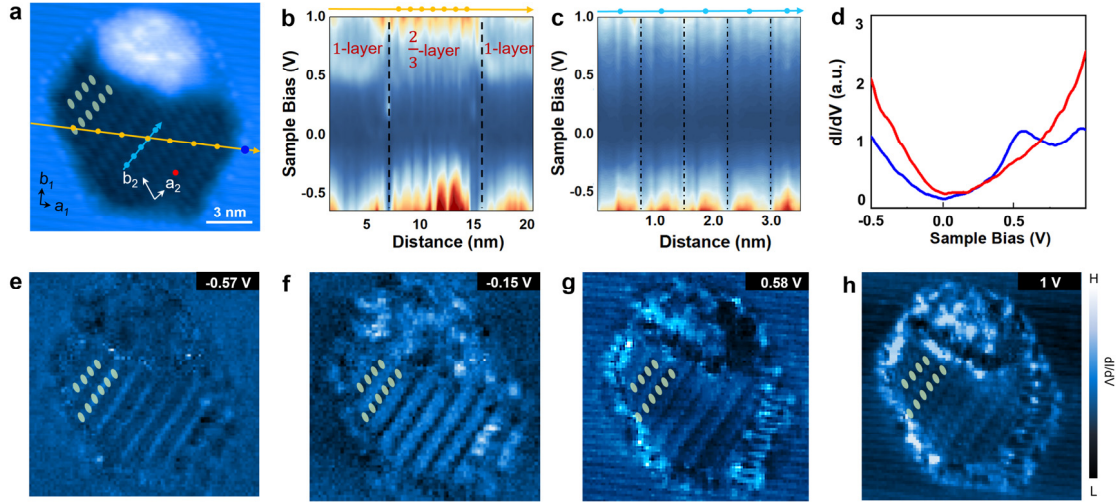

**Figure 4. Electronic properties of $\frac{2}{3}$-layer WTe$_2$. a,** Representative STM image of graphene-covered 1T′-WTe$_2$ after applying a STM tip pulse. **b,c,** Site-dependent STS spectra recorded along the yellow and blue arrows marked in panel a. **d,** Typical STS spectra acquired on graphene-covered 1T′-WTe$_2$ and $\frac{2}{3}$-layer WTe$_2$, respectively, as marked in panel a. **e-h,** Spectroscopic maps acquired at the same location as panel a under the sample bias of -0.57, -0.15, 0.58, and 1 eV, respectively.